\documentclass[onecolumn]{article}

\usepackage{amsmath}
\usepackage{amsfonts}
\usepackage{blindtext}
\usepackage[utf8]{inputenc}
\usepackage[T1]{fontenc}
\usepackage{graphicx}
\usepackage{float}
\usepackage{wrapfig}
\usepackage{url}
\usepackage{bm}
\usepackage{braket}
\usepackage{grffile}
\usepackage[usenames,dvipsnames]{xcolor}

\usepackage{natbib}

\newcommand{\tu}{\tilde{u}}
\newcommand{\tuu}{\tilde{u}^{(1)}}
\newcommand{\tud}{\tilde{u}^{(2)}}
\newcommand{\tulesu}{\tilde{u}^{(1), \rm LES}}
\newcommand{\tulesd}{\tilde{u}^{(2), \rm LES}}
\newcommand{\tup}{\tilde{u}'}
\newcommand{\tupu}{\tilde{u}'^{(1)}}
\newcommand{\tupd}{\tilde{u}'^{(2)}}

\newcommand{\xc}{x^{(0)}}

\newcommand{\bxc}{\bm{x}^{(0)}}

\newcommand{\tauu}{\tau^{(1)}}
\newcommand{\taud}{\tau^{(2)}}
\newcommand{\taulesu}{\tau^{(1), \rm LES}}
\newcommand{\taulesd}{\tau^{(2), \rm LES}}

\newcommand{\taupu}{\tau'^{(1)}}
\newcommand{\taupd}{\tau'^{(2)}}
\newcommand{\tS}{\tilde{S}}
\newcommand{\Su}{\tilde{S}^{(1)}}
\newcommand{\Sd}{\tilde{S}^{(2)}}

\newcommand{\Spu}{\tilde{S}'^{(1)}}
\newcommand{\Spd}{\tilde{S}'^{(2)}}
\newcommand{\tp}{\tilde{p}}
\newcommand{\tpu}{\tilde{p}^{(1)}}
\newcommand{\tpd}{\tilde{p}^{(2)}}
\newcommand{\retau}{\mathrm{Re}_\tau}

\begin{document}

\title{Two-point stress-strain rate correlation structure and\\
non-local eddy viscosity in turbulent flows}
\author{Patricio Clark Di Leoni, Tamer A. Zaki, George Karniadakis \&
Charles Meneveau}
\maketitle

\abstract{By analyzing the Karman-Howarth equation for filtered velocity fields in turbulent flows, we show that the two-point correlation between 
filtered strain-rate and subfilter stress tensors plays a central role in the evolution of filtered-velocity correlation functions. Two-point correlations-based {\it statistical priori tests} thus enable rigorous and physically meaningful studies of turbulence models. Using data from direct numerical simulations of isotropic and channel flow turbulence we show that local eddy viscosity models fail to exhibit the long tails observed in the real subfilter stress-strain rate correlation functions. Stronger non-local correlations may be achieved by defining the eddy-viscosity model based on fractional gradients of order $0<\alpha<1$ rather than the classical gradient corresponding to $\alpha=1$. Analyses of such correlation functions are presented for various orders of the fractional gradient operators. It is found that in isotropic turbulence fractional derivative order $\alpha \sim 0.5$ yields best results, while for channel flow $\alpha \sim 0.2$ yields better results for the correlations in the streamwise direction, even well into the core channel region.  In the spanwise direction, channel flow results show significantly more local interactions. The overall results confirm strong non-locality in the interactions between subfilter stresses and resolved-scale fluid deformation rates, but with non-trivial directional dependencies in non-isotropic flows. 
}


\section{Introduction}

Scale interactions in turbulent flows can be studied using the filtering approach, in which a spatial filter separates large from small scales \citep{Leonard74, Germano92, Pope}. Such studies are of particular interest in the context of  Large Eddy Simulations (LES) of turbulent flows \citep{Sagaut01, Pope}, where some but not all scales are solved for explicitly. The effect of scales smaller than the filter size is modeled by modifying the stress tensor in the equations. Most existing sub-filter (or subgrid-scale) models used
in practical LES today rely on the concept of eddy-viscosity, which
models the interaction between small and large scale turbulent structures in analogy to molecular viscosity. The sub-filter scale stress
tensor is set to be proportional to the strain rate tensor of the resolved
(filtered) motions at the same spatial position and time. While various models differ
in how the  proportionality factor, the eddy-viscosity, is specified (e.g. Smagorinsky \citep{Smagorinsky63}, dynamic Smagorinsky,  \citep{Germano91}, Vreman \citep{Vreman96}, WALE \citep{Nicoud99}), the approach is in essence a spatially and temporally {\it local} closure model.  

In recent years, several approaches to include non-local effects for
turbulent stresses have been proposed.  In the context of Reynolds
Averaged Navier-Stokes (RANS), non-local models for the Reynolds stresses and scalar fluxes have
been explored by \cite{Hamba95,Hamba04,Hamba05}, following the ideas of \cite{Kraichnan87}. These works derived explicit non-local in time and space expressions for the Reynolds stresses and scalar fluxes using Green's functions on the equations for the fluctuating velocity, then validated their results based on a-priori tests. More in general, closures with temporal memory arise from the Mori-Zwanzig formalism \citep{zwanzig2001nonequilibrium,li2017computing,parish2017non} for deriving evolution equations for coarse-grained dynamics. 
Spatially non-local expressions for the Reynolds stresses have also been obtained recently  
by the ``Macroscopic Forcing Method'' \citep{ShirianMani2019} or by treating turbulence dissipation as caused by singular spatiotemporal events interspersed in Euler equation evolution \citep{Pomeau19}.

Aiming to represent non-locality using compact
representations has led to consideration of fractional
operators to represent  fluxes and stress tensors.
Fractional differential operators can be roughly understood as operators
that when applied iteratively a certain number of times, coincide with a
given integer differential operator \citep{Samko, Lischke19}. When their order is not an integer, they can be understood as an operation lying
somewhere in between differentiation and integration and they are thus
inherently non-local.  Several definitions exist, each suited to
different problems.  Traditionally, fractional derivatives have been
used successfully to model anomalous diffusion and complicated materials
\citep{Carpinteri97, Caputo67}.  In turbulence, the application of non-local Levy walks to model intermittency \citep{Shlesinger87,Dubrulle98} has led to RANS models based on fractional Laplacians \citep{Chen06,Lischke19}. RANS modeling can also be achieved via other types of fractional operators
\citep{Egolf17,Epps18}.  In particular, recent developments of
channel flow modeling using the Caputo derivative \citep{Song18} to model the entire stress (viscous and Reynolds shear stress) show universal behavior of the fractional order as function of wall distance in viscous units.  For Large Eddy Simulations, a recent paper
\citep{Samiee19} proposes to model the subgrid-scale stress tensor using fractional derivatives motivated by considerations of non-Maxwellian (Levy-flight) equilibrium distributions of a Boltzmann equation. 

Data-driven approaches have also been on the rise in many areas of
science and in turbulence in particular \citep{duraisamy2019turbulence}.
Analyzing and predicting the performance of a subgrid-scale model {\it
a-priori} based on turbulence data requires careful consideration of
statistical measures of interest.  It is possible to establish several
statistical necessary and sufficient conditions that a sub-filter model for LES must satisfy  \citep{Meneveau94}. These conditions arise from analyzing balance equations for various statistical properties of the flow and
establishing how the sub-filter stress tensor affects the statistical
property of interest.  It is generally accepted that a most important
statistical feature of turbulent flow and LES is the mean kinetic energy
in the resolved flow.  Hence, a particularly important necessary
condition for a sub-filter  model states that the rate at which a
model extracts the kinetic energy from the large scales must be the same as
the rate of energy that is transferred from the large to the small
scales in the exact equations. Already  \cite{Lilly67} proposed an
energy dissipation balance condition to relate the Smagorinsky coefficient to the Kolmogorov constant and since then, satisfying the condition that a
subgrid model dissipate resolved kinetic energy at the correct rate lies at the heart of most eddy viscosity models. The rate of dissipation is a single-point statistical property. 

In this work we focus on basic two-point statistical features of turbulence. G.K. Batchelor's influential treatise ``{\it The Theory of Homogeneous Turbulence}'' \citep{batchelor1953theory} provides all the requisite conceptual background regarding the evolution of two-point statistics of velocity fluctuations in homogeneous turbulence and its various mathematical representations. Correctly capturing two-point correlations is of the utmost importance in turbulence modeling for LES, since these correlations and the concomitant energy spectral density describe the relative amplitudes of velocity fluctuations in the hierarchy of resolved structures in the flow simulated using LES. Motivated by the importance of two-point correlations,  in \S \ref{sec:twopoint} we formulate statistical conditions that subgrid scale models must satisfy regarding their two-point structure. In \S \ref{sec:initialcorr}, using direct numerical simulation (DNS) data from isotropic and channel flow turbulence, we examine such two-point structure and compare it with results from a canonical eddy-viscosity closure. 

Then, inspired by two-point statistically necessary conditions, we propose to include  non-local effects by relying on the compact expressivity of  fractional derivative operators in the definition of the eddy viscosity closures \ref{sec:nonlocal}. Using again DNS data from isotropic and channel flow turbulence, we  evaluate how well such non-local eddy viscosity closures can satisfy the two-point correlation statistical conditions mentioned above, as compared to the classical local versions \S \ref{sec:results}. Conclusions are provided in \S \ref{sec:conclusions}.    

\section{Two-point correlations of filtered velocity fields}
\label{sec:twopoint}

As summarized above, certain statistical conditions that an LES
subfilter model must satisfy can be derived by analyzing the evolution equations for the different order statistics of the fields, such as single-point moments $\langle \tu_i \rangle$,
$\langle \tu_i^2 \rangle$, multi-point moments, and so on \citep{Meneveau94}. We review
conditions based on  two point statistics as developed by
\cite{Meneveau94} and rephrase the results more conveniently in terms of the filtered
strain-rate tensor as opposed to the filtered velocity as was done in
\cite{Meneveau94}. We also generalize the prior derivations to the case
of non-homogeneous flow. Here we briefly summarize the  derivation of
the Karman-Howarth equations (the evolution equations for the two-point
velocity correlations) for the case of filtered velocity fields without
assuming homogeneity or isotropy, using the two-point method proposed
by \cite{Hill02}.

Starting from the Navier-Stokes equations for a divergence-free velocity field $\bm{u}$, the LES equations for the filtered fields $\tilde{\bm{u}} = \mathcal{F}(\bm{u})$, where $\mathcal{F}$ is a spatial
filtering operation, read as follows:

\begin{equation}
    \frac{\partial \tu_i}{\partial t} + \tu_k \frac{\partial \tu_i}{\partial x_k} =
    - \frac{\partial \tp}{\partial x_i} + \nu \frac{\partial^2
    \tu_i}{\partial x^2_k} - \frac{\partial \tau_{ki}}{\partial x_k}
    \label{nse_filt_idx}
\end{equation}
where
\begin{gather*}
    \tau_{ij} = \mathcal{F}(u_i u_j) - \mathcal{F}(u_i) \mathcal{F}(u_j)
    - \frac13 \left( \mathcal{F}(u_i u_i) - \mathcal{F}(u_i)^2 \right) \delta_{ij},
    \\
    \tilde{p}=\mathcal{F}(p)  
    + \frac13 \left( \mathcal{F}(u_i u_i) - \mathcal{F}(u_i)^2 \right),
\end{gather*}
are the deviatoric part of the subgrid-scale stresses and the modified pressure (where $p$ is the fluid pressure divided by the density), respectively.

Defining velocities at two points, $\tuu_i = \tu_i(\bm{x}^{(1)})$, $\tilde{u}_i^{(2)} =
\tilde{u}_.(\bm{x}^{(2)})$, and midpoint position \citep{Hill02}
$\bm{x}^{(0)}=(\bm{x}^{(2)}+\bm{x}^{(1)})/2$, where 
$\bm{x}^{(2)}=\bm{x}^{(1)}+\bm{r}$ and $\bm{r}$ is the displacement vector 
between the two points, one can multiply their corresponding evolution equations by 
each other,  sum them, rearrange and perform a statistical averaging 
operation (details are provided in the supplementary material). The result is the evolution equation for the velocity two-point correlation function $C_{uu}(\bm{r},\bm{x}^{(0)})=
\braket{\tuu_i \tud_i}$ 
\begin{equation}
    \frac{\partial C_{uu}}{\partial t}
    = - T(\bm{r},\bm{x}^{(0)}) +
    P_0(\bm{r},\bm{x}^{(0)})+V(\bm{r},\bm{x}^{(0)}) - \frac{\partial
    }{\partial \xc_k} \langle \tud_i \tauu_{ki} +\tuu_i
    \taud_{ki}\rangle - \langle \tauu_{ki} \Sd_{ki} + \taud_{ki}
    \Su_{ki} \rangle,
\label{kheq_general}
\end{equation}
where

\begin{equation}
    \tilde{S}_{ik} = \frac12 \left( \frac{\partial \tu_i}{\partial x_k} + \frac{\partial \tu_k}{\partial x_i} \right)
    \label{filt_strain_rate}
\end{equation}
is the filtered strain-rate and
\begin{equation*}
    T(\bm{r},\bm{x}^{(0)}) = \frac{\partial  }{\partial r_k} \left(
    \langle \tuu_i \tud_i \tud_k\rangle -  \langle\tuu_i \tud_i \tuu_k
    \rangle \right) - \frac12 \frac{\partial  }{\partial \xc_k} \left(
    \langle \tuu_i \tud_i \tud_k\rangle  +   \langle \tuu_i \tud_i
    \tuu_k \rangle  \right), 
\end{equation*}
\begin{equation*}
    P_0(\bm{r},\bm{x}^{(0)})  =  - \frac{\partial }{\partial \xc_i}
    \left( \langle \tuu_i \tpd \rangle + \langle\tud_i \tpu,
    \rangle\right)
    \,\,\,\,  {\rm and} \,\,\,\,\,
    V(\bm{r},\bm{x}^{(0)}) = 2 \nu \frac{\partial^2  C_{uu}}{\partial r_k r_k} + \frac12 \nu \frac{\partial^2
   C_{uu}}{\partial \xc_k \xc_k}.
\end{equation*}

This equation holds for velocity fields  $\tu_i$ obtained from first
solving Navier-Stokes equations and then filtering the results, and also
for velocity fields  $\tu^{\rm LES}_i$ arising from solving LES
equations in which the SGS stresses are replaced by a subgrid model,
i.e. $\tau_{ij}^{\rm LES}$. For LES and filtered Navier-Stokes to yield the same two point statistical moment evolution of $\braket{\tuu_i \tud_i}$ as
well as the two-point third-order moments  $\langle \tuu_i \tud_i
\tud_k\rangle$ requires as a statistically necessary condition \citep{Meneveau94} that  

\begin{align}
    \langle \taulesu_{ki} \Sd_{ki} \rangle + \langle \taulesd_{ki} \Su_{ki} \rangle
    &=
    \langle \tauu_{ki} \Sd_{ki}  \rangle + \langle \taud_{ki} \Su_{ki} \rangle  \,\,\,\,\,\,\,\, {\rm and}  \label{eq:dissip1}
    \\
    \nonumber \\
    \frac{\partial }{\partial \xc_k} \left(\langle \tulesd_i \taulesu_{ki} \rangle
    +\langle \tulesu_i
    \taulesd_{ki}\rangle \right)
    &=
    \frac{\partial }{\partial \xc_k} \left(\langle \tud_i \tauu_{ki}
    \rangle + \langle \tuu_i \tauu_{ki}\rangle \right) \label{eq:dissip2}
\end{align}


If the flow is spatially homogeneous  and isotropic, all derivatives with respect
to $\bxc$ vanish and equation \eqref{eq:dissip2} is irrelevant while  the last term in 
equation \eqref{eq:dissip1} simply becomes $2\langle \tauu_{ki}
\Sd_{ki}\rangle$. In addition, in isotropic flow terms only depend on $|{\bf r}|$
(also, tensor contractions may be simplified in terms of single components, but these consequences will not be utilized explicitly here since data from DNS will be used for which all components are available). Equation~\eqref{kheq_general} then becomes

\begin{equation} 
   \frac{1}{2} \frac{\partial C_{uu}(r)}{\partial t}
    = 
    -   \frac{\partial  }{\partial r_k}
   \langle \tuu_i \tud_i \tud_k\rangle  
    +   \nu \frac{\partial^2  C_{uu}(r)}{\partial r_k r_k}
    -   \langle \tau_{ki}({\bm x}) \tilde{S}_{ki}({\bm x}+{\bm r}) \rangle, 
\label{kheq_hit}
\end{equation}
so that a necessary condition for LES to correctly predict two-point
moments of the resolved field reduces to 
\begin{equation}
    \langle \tau^{\rm LES}_{ki}({\bm x}) \tilde{S}^{\rm LES}_{ki}({\bm x}+{\bm r}) \rangle = \langle \tau_{ki}({\bm x}) \tilde{S}_{ki}({\bm x}+{\bm r}) \rangle. 
    \label{kheq_simplest}
\end{equation}
For the case of kinetic energy, i.e. for the single point case ${\rm r}
= 0$, the familiar condition is recovered where LES
should correctly predict the SGS rate of dissipation, i.e. $ \langle
\tau_{ki}^{\rm LES} \tS^{\rm LES}_{ki} \rangle = \langle \tau_{ki}
\tS_{ki} \rangle $. The Fourier transformed version of this expression
(involving 
$\langle \hat{\tau}_{ki}({\bf k}) \hat{\tilde{S}}^*_{ki}({\bf k})\rangle $,
where a {\it hat} denotes 3D Fourier transform and ${\bf k}$ is the wavenumber vector) was used by \cite{Cerutti00} to measure spectral eddy-viscosity distributions. In the present work we focus attention on physical space descriptions to highlight the strength of spatial correlation at various distances.

 
The case of channel flow  is statistically homogeneous  in  the  two wall-parallel
directions but inhomogeneous in the wall normal direction. And, the presence
of walls and a mean pressure gradient break isotropy.   Due to the
special importance of mean shear to this flow, it is convenient to
separate the resolved flow also into its statistical mean and
fluctuating variables according to $    \tu_i = \langle  \tu_i
\rangle + \tu_i' $. In channel flow, taking the $x_1$ direction to be streamwise
and $x_2$ to be wall normal, the averaged flow variables do not depend
on $x_1$, $x_3$, 
and  $\langle \tu_2 \rangle = \langle
\tu_3 \rangle = 0$. Taking these facts into consideration and
restricting the displacement between the two points in the correlations
to be in the horizontal direction (i.e. ${\rm r} = (r_1,0,r_3)$),
equation~\eqref{kheq_general} becomes

\begin{gather}
\begin{aligned}
     \frac{\partial \langle \tupu_i \tupd_i \rangle}{\partial t} 
    =& - T'(\bm{r},\bm{x}^{(0)}) + P_0'(\bm{r},\bm{x}^{(0)})
    + V'(\bm{r},\bm{x}^{(0)}) + \mathcal{P}(\bm{r},\bm{x}^{(0)})
    \\
    &- \frac{\partial }{\partial \xc_2} \langle \tupd_i \taupu_{2i} +
    \tupu_i \taupd_{2i}\rangle
    - \langle \taupu_{ki} \Spd_{ki} + \taupd_{ki} \Spu_{ki} \rangle,
\end{aligned}
\label{eq:correl}
\end{gather}

where
\begin{equation*}
    \mathcal{P}(\bm{r},\bm{x}^{(0)}) =
    \braket{\tupd_1 \tupu_2} \frac{\partial \braket{\tup_1}}{\partial
    \xc_2}
\end{equation*}
and the primed terms $T'$, $P_0'$ and $V'$ correspond to the previously
defined $T$, $P_0$ and $V$ but for the primed fields, respectively. We note that the last two terms in equation \eqref{eq:correl} may also be combined and re-expressed
in terms of the total variables (without decomposing into fluctuating
and mean quantities) if the displacement is taken in the horizontal
plane, since then $\langle \tilde{S}_{ki} \rangle$ and $ \langle
\tau_{ki} \rangle$ are the same at locations (1) and (2).  Therefore,
again we can state that a necessary condition for LES to generate each
of the terms involving filtered velocities and mean velocities in the
above equation accurately requires that the equality in 
\eqref{kheq_simplest}  must hold. 

\section{Stress-strain rate correlations in isotropic and channel flow data}
\label{sec:initialcorr}

Having established the relevance of the stress-strain rate correlation
function for understanding interactions between scales in turbulence, we examine such correlations in two canonical
data sets and compare the results to the correlations arising from
classic local eddy-viscosity models. We first start by describing the datasets, then the way in which we process the data, and finally measure and report the aforementioned correlation functions.

The homogeneous and isotropic turbulence data come from a simulation of the Navier-Stokes equations
including a forcing term performed on a periodic grid of $1024^3$
grid points using a pseudo-spectral parallel code \citep{Lietal2008}. The viscous term was
integrated analytically using an integrating factor, while all other
terms were integrated using a second order Adams-Bashforth scheme. A
combination of phase-shift and $2\sqrt{2}/3$ truncation was used to
de-alias the simulation. The forcing term is such that kinetic energy in modes
with wavenumber less or equal to 2 was kept constant.
The Kolomogorov length $\eta \sim (\epsilon^3/\nu)^{1/4}$, where $\epsilon$ is the mean energy transfer rate and $\nu$ the molecular kinematic viscosity was about half of the grid spacing. The averaged Taylor-scale Reynolds number of the simulation is $\mathrm{Re}_\lambda \sim 433$. The data are available on the public JHTDB database server \cite[for more information, see][]{Lietal2008}. 
For our analysis, we use data from eleven independent snapshots distributed over about five large-eddy turnover times. All two-point
correlations were performed along a given Cartesian direction and correlation functions were then averaged over the three Cartesian directions and over the 11 snapshots in time.

The channel flow data come from two  different  
friction Reynolds numbers, $\mathrm{Re}_\tau \sim 1{,}000$ and  
$\mathrm{Re}_\tau \sim 5{,}200$. Both simulations solve the Navier-Stokes
equations in a domain with periodic boundary conditions in two
directions (the horizontal directions parallel to the walls) and no-slip boundary conditions
in the other direction (the vertical direction). The data are also available at JHTDB.  For more detailed information regarding both channel flow datasets, see  \cite{Graham16} and \cite{Lee15}, respectively.

We filter the velocity using a top-hat box filter with different filter lengths. Results using different types of filters (Gaussian and spectral) are also presented although the focus will be on results from the most spatially local filter (box filter). For the case of the channel flows, the filtering is performed only in the horizontal directions. The true subgrid-scale stresses (their deviatoric part) are calculated using their usual definition

\begin{equation}
    \tau_{ij} = \widetilde{u_i u_j} - \tilde{u}_i \tilde{u}_j - \frac{1}{3}( \widetilde{u_k u_k} - \tilde{u_k} \tilde{u_k}) \delta_{ij},
\end{equation}
while the filtered strain-rates are calculated using Eq.~\eqref{filt_strain_rate} and where the derivatives are calculated using a second order centered finite difference scheme.

When considering predictions from eddy-viscosity models, in order to avoid introducing model-dependent effects from various possible choices of eddy viscosity (e.g. Smagorinsky \citep{Smagorinsky63}, dynamic Smagorinsky,  \citep{Germano91}, Vreman \citep{Vreman96}, WALE \citep{Nicoud99}), we focus on the simplest version, namely a ``constant eddy viscosity'' model. Specifically,

\begin{equation}
    \tau^{\rm LES}_{ij} = -2 \nu_{\rm LES} \, \tilde{S}_{ij}
\end{equation}
with $\nu_{\rm LES}$ a constant (which in our analysis will ultimately not play a role due to the normalization to be used). The averaging operation is performed in the three Cartesian spatial directions and over time snapshots when analyzing isotropic turbulence data, while for the case of channel flow no averaging is  performed in the vertical direction.  Then, the two-point correlations between either the true subgrid-scale stresses and the filtered strain-rates, and the modeled sub-grid stress and the strain rates are obtained by averaging the product of the displaced fields and normalizing to unity at $r=0$. In other words,  we normalize each correlation function by its own rate of subgrid-scale dissipation rate so as to focus on the spatial correlation structure independent of the mean dissipation rate (hence rendering the value of $\nu_{\rm LES}$ irrelevant for our analysis).

\begin{figure}
    \centering
    \includegraphics[width=0.45\textwidth]{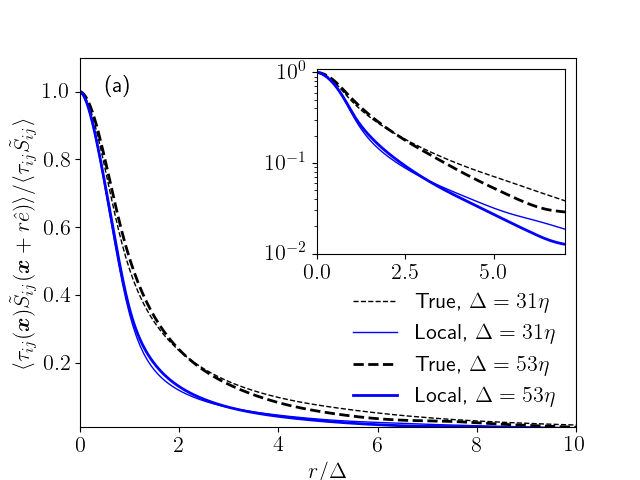}
    \includegraphics[width=0.45\textwidth]{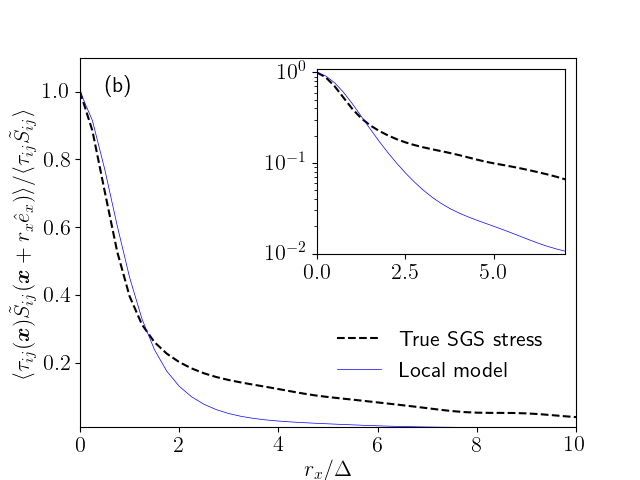}
    \caption{Two-point correlations between different subgrid-stresses
    and the filtered strain rates in (a) homogeneous isotropic turbulence and (b) channel flow at $Re_\tau = 1000$ at $y^+ = 90$.  Data is filtered using a top-hat filter at two different filter
    sizes and in all three directions for the homogeneous isotropic case (a), while only one filter size and horizontal filtering was used for the channel flow case (b), at $Delta^+=49$ (or $\Delta/h=0.049$).}
    \label{motivation}
\end{figure}

In Fig.~\ref{motivation}(a) we show the two-point correlations using both the true subgrid-scale stress tensor and the modeled one extracted from the homogeneous isotropic turbulence data using two different filter sizes $\Delta=31\eta$ and $\Delta=53\eta$ (the figures' insets show the same plots but in semilog scale to better visualize the long-distance tails). While close to the origin both the real and the local eddy-viscosity cases behave similarly, the correlation function for the real sub-filter stress sustain correlations with the strain-rate for longer distances than the case of the local eddy-viscosity. A similar result is seen in Fig. ~\ref{motivation}(b) where such correlation functions are shown for the $Re_\tau=1,000$ channel flow at some height in the logarithmic region $y^+ = 1000$, for two points separated along the streamwise direction. 
 
As we have seen before, if one wishes an LES to provide realistic predictions of filtered velocity correlation functions (or spectra) down to distances as close as possible to the filtering scale $\Delta$,  a model should be such that the two-point correlation between subfilter-stress and strain-rate are correctly reproduced during a-priori data analysis. Present results demonstrate that local models cannot correctly capture the relatively long tails in these two-point correlations. Correlations between distances $r=\Delta$ to about $5\Delta$ are underestimated by local models, yielding only about 50\% of the real correlation. 
Interestingly, we recall that it was already observed that the correlations between sub-grid stresses and velocity increments also decay faster in LES than in the true cases \citep{Linkmann18}. It is important to emphasize that since we are normalizing the correlations by their value at the origin, the dependence on any constant scalar prefactor is eliminated. We have tested that even if the scalar prefactor is position dependent, as it is with various variants of the eddy-viscosity model (e.g. classical Smagorinky model), the normalized correlations remain very similar to our present results and thus are not presented here.

\begin{figure}
    \centering
    \includegraphics[width=0.7\textwidth]{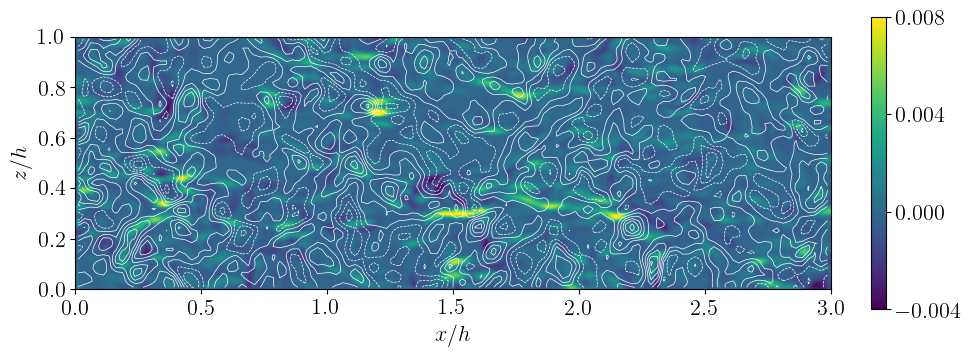}
    \caption{Contour plot of streamwise normal deviatoric subgrid-scale stress $\tau_{11}$ (filled colours, in simulation units) superimposed with contour lines of $-\tilde{S}_{11}$ on a horizontal plane of the channel flow at at $y^+=90$. The filtering scale is $\Delta^+=49$ ($\Delta/h = 0.049$). The maximum contour line of $-\tilde{S}_{11}$ shown is at $0.36$ (in simulation units), all other have an equal spacing of $0.06$, dashed lines indicate negative values.}
    \label{tau_viz}
\end{figure}

Fig.~\ref{tau_viz} shows a sample contour plot of $\tau_{11}$ (filled colours) superimposed with a contour plot (lines) of $-\tilde{S}_{11}$ on a horizontal plane at $y^+=90$ of the channel flow data at $Re_\tau=1000$. The elongated features of the real stress are apparent, extending over distances far exceeding the filter scale ($\Delta/h \sim 0.05$). The features of $-\tilde{S}_{11}$ appear more isotropic with less `non-locality' in the $x$-direction compared to $\tau_{11}$. As is well known, pointwise comparisons between instantaneous distributions of real and modeled subgrid-scale stresses using variants of eddy-viscosity models typically lead to very low correlation coefficients (typically less than 20\%). But such instantaneous {\it a-priori tests} have little prognostic power regarding the statistics resulting from LES. Hence, we continue instead to focus on the two-point correlations ({\it statistical a-priori tests} for which the interpretation is clear, following the discussion in \S \ref{sec:twopoint}.

\section{Non-local eddy viscosity modeling}
\label{sec:nonlocal}

\subsection{Motivation for non-local eddy viscosity modeling}

It is instructive to begin by discussing kinetic energy
dissipation. One possible motivation for the classic local eddy
viscosity modeling in LES is provided directly by the condition of
matching SGS energy dissipation rates. Specifically, suppose we wish to
ensure that energy will be dissipated at some `true' rate   $-\langle
\tau_{ki} \tS_{ki} \rangle $. The LES will generate fluctuations including fluctuating filtered strain rates. Thus  one can be sure that its variance, i.e. the quantity $ \langle \tS^{\rm LES}_{ki} \tS^{\rm LES}_{ki} \rangle$, will be positive. Its magnitude will depend on the fluctuation amplitudes of $\tS^{\rm LES}_{ki}$ but will not involve any subtle cancellations of oppositely signed values. Hence, if one sets the subgrid-scale stresses
$\tau_{ki}^{\rm LES}$ (where we are always working with the traceless tensor) proportional to  $-\tS^{\rm LES}_{ki}$, i.e.
$\tau_{ki}^{\rm LES} \sim - \tS^{\rm LES}_{ki}$, one will be guaranteed a mean dissipation rate that will be proportional to the nonzero $\langle \tS^{\rm LES}_{ki} \tS^{\rm LES}_{ki} \rangle$  value resulting in LES. The actual value can be controlled by choices of  SGS eddy viscosity, as in the standard model $\tau_{ki}^{\rm LES}=-2 \nu_{\rm LES} \tS^{\rm LES}_{ki}$.
 
Now, we wish to generalize this statement to the case of ensuring that
two-point moment between the subfilter stress and the filtered strain-rate tensor at some particular displaced position ${\bf r}^\prime$ are predicted correctly.  A possible way to guarantee that the two point correlation $\langle \tau^{\rm LES}_{ki}({\bf x}) \tS^{\rm LES}_{ki}({\bf x}+{\bf r}^\prime) \rangle$ is non-zero with its magnitude set by some prefactor, is to select  $\tau^{\rm
LES}_{ki}({\bf x})$ to be proportional not to the local value of the
filtered strain rate, but to $ \tS^{\rm LES}_{ki}({\bf x}+{\bf
r}^\prime)$ at the desired point, i.e. $\tau^{\rm LES}_{ki}({\bf x}) \sim - \tS^{\rm LES}_{ki}({\bf x}+{\bf r}^\prime)$. In general we will want to enforce such a condition for all possible ${\bf r}^\prime$, and so a weighted superposition of strain-rates at different locations can be envisioned:

\begin{equation}
    \tau^{\rm LES}_{ki}({\bf x}) = - \int K({\bf r}^\prime) \,
    \tS^{\rm LES}_{ki}({\bf x}+{\bf r}^\prime) \,\,d^3 {\bf r}^\prime,
    \label{eq:nonlocal1}
\end{equation}
where $K({\bf r}^\prime)$ represents an eddy-viscosity appropriate for displacement ${\bf r}^\prime$.

Multiplying Eq. \eqref{eq:nonlocal1} by $\tS^{\rm LES}_{ki}({\bf x}+{\bf r})$ and ensemble averaging yields the two-point correlation relevant to   correctly predicting two-point velocity correlations. The result, for homogeneous turbulence, can be written as 
\begin{equation}
    -\langle \tau^{\rm LES}_{ki}({\bf x})  \tS^{\rm LES}_{ki}({\bf
    x}+{\bf r}) \rangle  =  \int K({\bf r}^\prime) \,  \langle  \tS^{\rm
    LES}_{ki}({\bf x}+{\bf r}) \tS^{\rm LES}_{ki}({\bf x}+{\bf
    r}-{\bf r}^\prime) \rangle  \,\, d^3 {\bf r}^\prime,
\end{equation}
i.e. a convolution between a kernel and the strain-rate two-point
correlation function. Assuming that the latter decays as function of
displacement differently than the true correlation $ \langle \tau^{\rm
d}_{ki}({\bf x})  \tS_{ki}({\bf x}+{\bf r}) \rangle$ (as Fig. \ref{motivation} shows occurs in turbulence), then the
convolution with a kernel $K({\bf r}^\prime)$ enables one to generate a
SGS model that may display improved two-point correlations between
stress and strain-rate.   

Many options for the kernel $K({\bf r}^\prime)$ could be envisioned, and many ways to optimally find it from data can be developed.  Here we
propose to explore the applications of fractional calculus since such
operators enable compact representations and we may express the model
without introducing length-scales a-priori into the problem (as we shall see later on, effectively we will be introducing modeling length-scales anyhow, but at a later stage).  In the next section, we set the stage for definitions of fractional gradients that can be applied in 3D to gradient vector fields.

\subsection{Fractional gradient based non-local eddy-viscosity} 

Most applications of fractional derivatives are essentially one-dimensional, e.g. in time to represent memory effects, for spatially 1D problems, or using fractional Laplacians that do not discriminate between different directions. Recent efforts to develop fractional vector calculus and directionally dependent gradient operators include \cite{Meerschaert06, Tarasov08}.  Here, we use a multidimensional generalization of the  
Caputo fractional derivative \citep{Caputo67, Samko}. In 1D, the Caputo fractional derivative usually takes the form 
\begin{equation}
    {}_C D^\alpha_{0,x} f(x) = \frac{1}{\Gamma(1-\alpha)} \int^x_0
    \frac{\frac{df}{dx} (x')}{(x-x')^\alpha} dx'.
    \label{caputo1d}
\end{equation}
While this non-symmetric definition, where the limits of integration go
from 0 (or some other finite limit) to the point of evaluation is useful for many problems, and has been used in non-local closures of channel flow RANS \citep{Song18}, it is not applicable to flows in 3D where there may be various important directions. The need to constrain the domain of integration also arises, as in practice integrating over the whole physical domain would be prohibitively expensive. A symmetrized and truncated version of the Caputo derivative may be written as:
\begin{equation}
    {}_{\rm symC} D^\alpha_{R} f(x) = \frac{1}{2\Gamma(1-\alpha)}
    \int^{x+R}_{x-R}
    \frac{\frac{df}{dx} (x')}{|x-x'|^\alpha} dx'.
    \label{symm_caputo}
\end{equation}

Next, a definition of a  vector gradient is required. Some definitions of fractional gradients resort to just taking a 1D fractional derivative along each dimension \citep{Meerschaert06, Tarasov08}. Such a definition would, however, not be useful in general, as the gradient operation would not be invariant under arbitrary rotations of the coordinate system. Instead, we keep the directionally sensitive derivative inside the integral in one direction (as in the Caputo derivative), but integrate in all directions over a ball of radius $R$ \cite{Caputo15}, according to:

\begin{equation}
    D^\alpha_i u_j (\bm{r}) = \frac{1}{\Omega_d \Gamma(1-\alpha)}
    \int_{|\bm{r}-\bm{r}'|\leq R} \frac{ \frac{\partial u_j}{\partial
    x_i} (\bm{r}')}{ |\bm{r} - \bm{r}'|^{\alpha+d-1}} d^d \bm{r}'
\end{equation}
where $\Omega_d$ is the $d$-dimensional solid angle. In three dimensions (3D), this definition becomes

\begin{equation}
    D^\alpha_i u_j (\bm{r}) = \frac{1}{4 \pi \Gamma(1-\alpha)}
    \int_{|\bm{r}-\bm{r}'|\leq R} \frac{ \frac{\partial u_j}{\partial
    x_i} (\bm{r}')}{ |\bm{r} - \bm{r}'|^{\alpha+2}} d^3 \bm{r}'.
    \label{eq:fractionalgradient}
\end{equation}
The result depends also on the radius $R$ which in practice will be chosen ``large enough'' to capture non-locality and generate results that do not depend strongly on $R$.  

It is possible to show by performing integration by parts (following \cite{Li07})
that, as long as the field $u$ has a well defined second derivative, this definition complies with the following limiting behavior at $\alpha$ approaching unity from below:

\begin{equation}
D^\alpha_i u_j (\bm{r}) \xrightarrow[\alpha \to 1^-]{}
\frac{\partial u_j}{\partial r_i} (\bm{r}).
\end{equation}
That is to say, traditional gradient operation corresponds to $\alpha = 1$. 

In the limit of $\alpha \to 0^+$, we obtain
\begin{equation}
D^\alpha_i u_j (\bm{r}) \xrightarrow[\alpha \to 0^+]{}
\frac{1}{4\pi} \int^\pi_0 \int^{2\pi}_0 \int^R_0 \frac{\partial
u_j}{\partial r_i}(\bm{r} - \tilde{\bm{r}}) d\tilde{r} \sin \theta
d\theta d\phi,
\end{equation}
(when expressed in spherical coordinate system). For example, for a spatially constant velocity gradient within the sphere of radius $R$, the gradient in the limit $\alpha \to 0^+$ becomes:
\begin{equation}
D^\alpha_i u_j (\bm{r}) \xrightarrow[\alpha \to 0^+]{}
R \, \frac{\partial
u_j}{\partial r_i} ,
\end{equation}
similar to a velocity increment (structure function) over a distance $R$.

The units of the fractional gradient are velocity divided by (length)$^\alpha$. 
In a turbulent flow with weak mean gradients, and as long as $\alpha > 0$, the definition of the derivative is expected to converge for sufficiently large $R$, as contributions from different directions will mostly cancel. But possible dependencies on $R$ will be examined quantitatively during the analysis since a-priori some dependence on $R$ cannot be excluded.  

Now that we have defined a fractional gradient operator, we can also define the symmetric part, i.e. the  fractional strain-rate tensor, according to

\begin{equation}
    \tS^\alpha_{ij} = \frac{1}{2} \left(D^\alpha_i \tu_j+D^\alpha_j \tu_i \right).
\end{equation}

\begin{figure}
    \centering
    \includegraphics[width=0.45\textwidth]{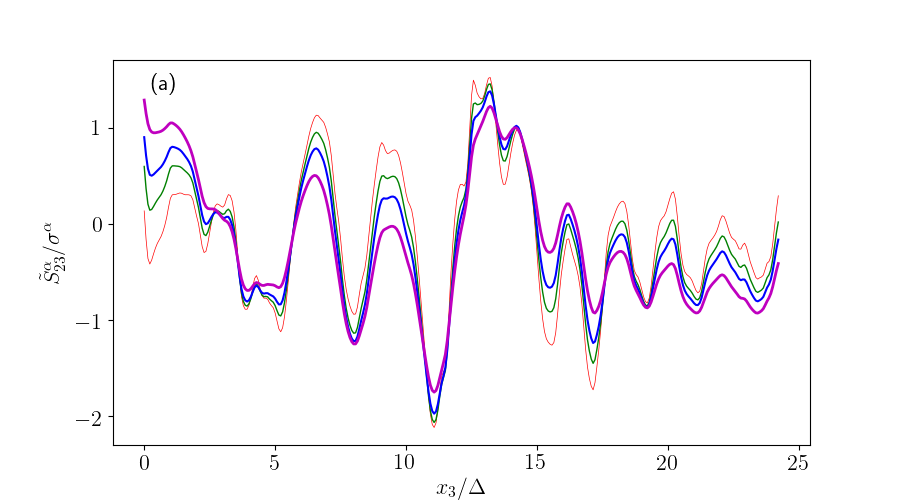}
    \includegraphics[width=0.45\textwidth]{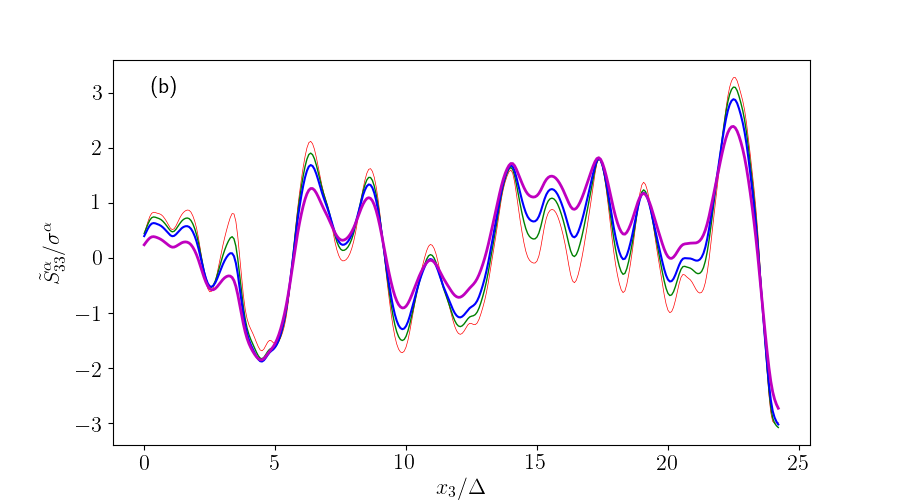}
    \caption{Effects of $\alpha$: Representative signal of two components of the fractional strain rate tensor $\tS_{ij}^{\alpha} = \frac{1}{2}(D^\alpha_i \tu_j+D^\alpha_j \tu_i)$
    normalized by its respective standard deviation for different
    fractional orders. Data are from DNS of isotropic turbulence at $Re_\lambda \sim 433$. (a) is for $ij=23$, while (b) is for $ij=33$. The curves correspond to $\alpha = 1, 0.7, 0.5$ and 0.2, from light to dark color, respectively.}
    \label{strains}
\end{figure}

In order to provide qualitative insights regarding the fractional gradient, we apply the gradient operator to filtered velocity fields from the isotropic turbulence data from DNS described before (at $\Delta = 31 \eta$). Details on the numerical calculation of the fractional strain-rate are presented in the Appendix~\ref{numerical_method}. Sample signals of transverse $\tilde{S}^\alpha_{23}$ and normal $\tilde{S}^\alpha_{33}$ velocity gradient tensor elements across parts of the computational domain are shown in  Fig.~\ref{strains}. Each of the curves are normalized
by their respective standard deviations $\sigma_\alpha$. 
The standard gradient tensor signals ($\alpha = 1$) are shown as the light curve.  The signals corresponding to lower values of $\alpha$ display smaller excursions in general, consistent with the idea that they are more non-local.  It is interesting to note that  even if subtle, the most non-local case ($\alpha=0.2$) still retains significant small scale structure in the signal (at the filter scale $\Delta$) even though it is the most non-local case considered. 

\begin{figure}
    \centering
    \includegraphics[width=0.45\textwidth]{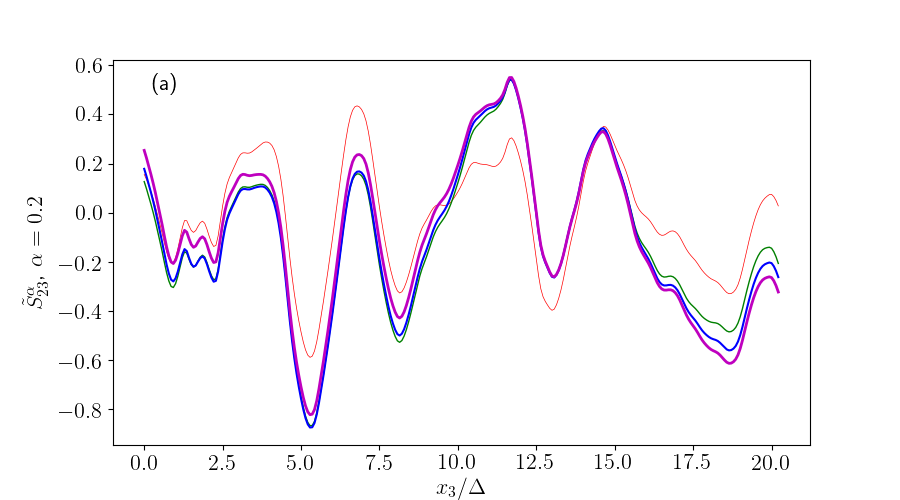}
    \includegraphics[width=0.45\textwidth]{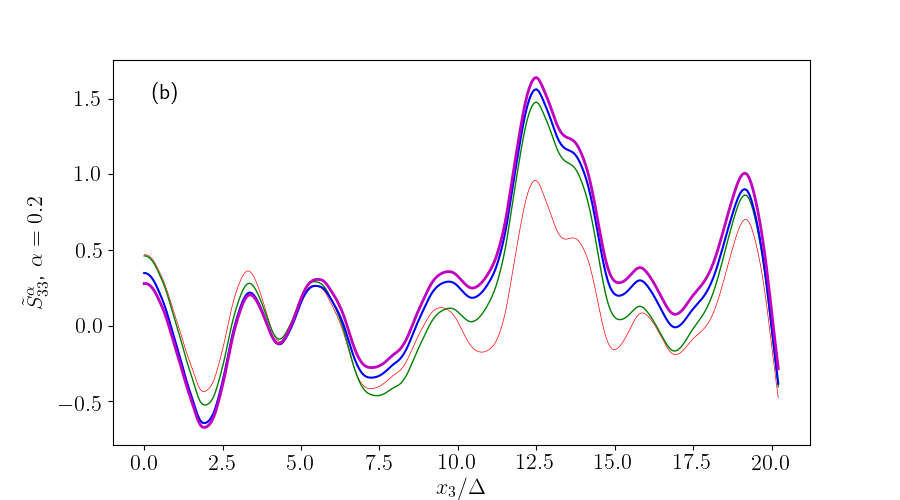}
    \caption{Effects of $R$:  Two components of the fractional strain rate tensor $\tS_{ij}^{\alpha} = \frac{1}{2}(D^\alpha_i \tu_j+D^\alpha_j \tu_i)$
    for the same
    fractional order $\alpha=0.2$, but different cut-off radius $R$. (a) is for $ij=23$, while (b) is for $ij=33$ The different curves correspond to $R = 1\Delta, 3\Delta, 5\Delta$ and $7\Delta$, from light to dark color, respectively.}
    \label{strains_R}
\end{figure}

In Fig.~\ref{strains_R} we present plots of the same two components of the $\tilde{S}^\alpha_{ij}$ for a fixed value of $\alpha=0.2$ but calculated using different cut-off radius $R$. As mentioned earlier, the fractional derivative \eqref{eq:fractionalgradient} is not independent of $R$, but we can expect results from turbulent flow with weak mean gradients to vary less and less as $R$ increases. The strain rate signals shown in Fig.~\ref{strains_R} are consistent with such  behavior, with relatively small sensitivity to $R$ for $R\geq 5\Delta$ in these examples. Note that these results are for a small value of $\alpha$ (0.2). For larger values of $\alpha$ the sensitivity to $R$ is less marked.

Based on the fractional strain-rate tensor, we may now define a fractional eddy-viscosity closure for the deviatoric part of the subgrid-scale stress, according to
\begin{equation}
    \tau^\alpha_{ij} = -2 \nu_\alpha \tS^\alpha_{ij},
\end{equation}
where $\nu_\alpha$ is the $\alpha$-dependent subgrid scale eddy viscosity, having units of velocity times (length)$^\alpha$. For the purposes of data analysis in this paper, similarly to what we showed in the previous section for local models, we will take $\nu_\alpha$ to be constant. Practical applications of such models will of course require specification of $\nu_\alpha$, which would indirectly involve specification of a length-scale (e.g. setting $\nu_\alpha \sim \Delta^{2\alpha} |\tS^\alpha|$). 

The form based on the fractional gradient defined as in Eq. \eqref{eq:fractionalgradient} has the following desirable properties: It is Galilean invariant, it is rotationally invariant, and the stress enters as a divergence in the momentum equation so that it obeys a traditional Gauss theorem (i.e. its volume integral only leaves surface fluxes). 

\section{Results} 
\label{sec:results}

In this section, we test the effectiveness of fractional-gradient based eddy-viscosity modeling to reproduce the desired two-point correlation structure of subgrid stresses and filtered strain-rate tensors in isotropic and channel flow turbulence.
\subsection{Homogeneous and isotropic turbulence} 

\begin{figure}
    \centering
    \includegraphics[width=0.4\textwidth]{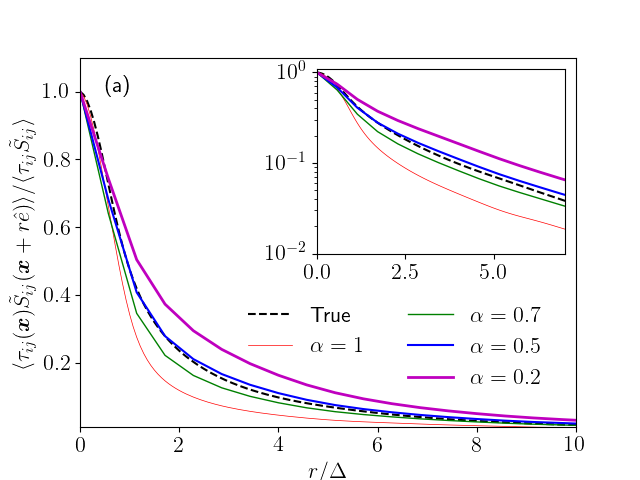}
    \includegraphics[width=0.4\textwidth]{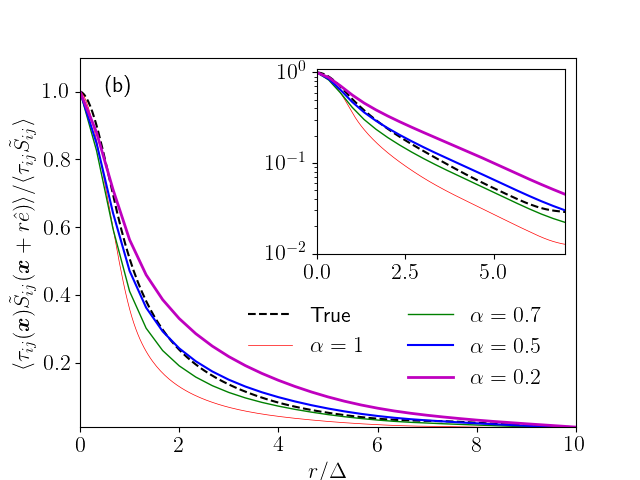}
    \caption{Two-point correlations between different subgrid-stresses
    and the fractional filtered strain rates of different orders in homogeneous isotropic turbulence.
    Results are for a top-hat filter at $\Delta = 31\eta$ (a), and  $\Delta = 53\eta$ (b).  In both we use $R=5\Delta$}
    \label{HIT}
\end{figure}

Figure~\ref{HIT} shows the two-point correlation between the filtered
strain rates and the different SGS stresses: the true one coming from the DNS (dashed line), the one modeled by the traditional local eddy-viscosity model ($\alpha=1$), and three cases using the fractional gradient based model at
different fractional orders (but same cut-off radius $R=5\Delta$) for two different filter sizes in the inertial range of turbulence. As can be seen, the fractional models generate longer correlations than the local model. The fractional order of $\alpha = 0.5$ reproduces the degree of non-locality found in the DNS case well, for both filter scales analyzed.  As was observed in 
Fig.~\ref{motivation}, the decay of correlations seems to scale with $r/\Delta$ also for the fractional models.

\begin{figure*}
    \centering
    \includegraphics[width=0.4\textwidth]{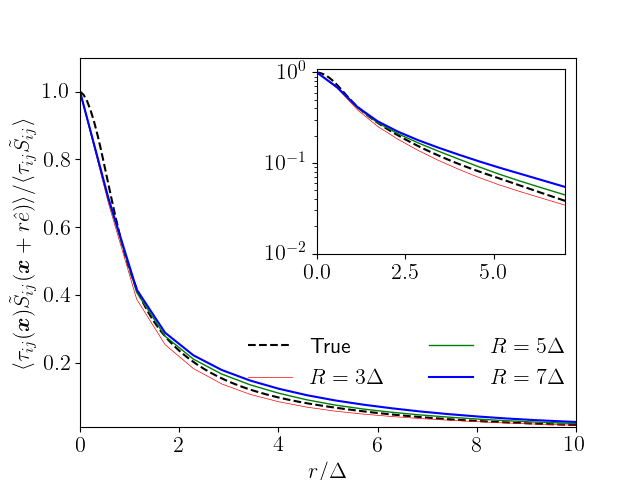}
    \caption{Two-point correlations between different subgrid-stresses
    and the filtered strain rates in homogeneous isotropic turbulence. The modeled stresses were all calculated using $\alpha=0.5$, but different values of $R$.}
    \label{HIT_R}
\end{figure*}

Figure~\ref{HIT_R} presents the two-point correlations functions for the true case and for the fractional case using the fractional order  $\alpha=0.5$ and same filter size and type, but using different cut-off radii $R$. As expected from the results shown in Fig.~\ref{strains_R}, the behaviour of the fractional models does depend slightly on $R$. When using a smaller cut-off radius, the models behave moderately more ``locally'' for the same fractional order, as less non-local information is used. This indicates that when fine-tuning practical applications of fractional models, both the order and the domain of integration will have to be considered. Differences in the correlations produced by the different parameters do appear to get smaller the larger the cut-off radius, again suggesting that although the fractional derivatives do not formally converge for any arbitrary field, they might do so in a turbulent flow.

\begin{figure*}
    \centering
    \includegraphics[width=0.4\textwidth]{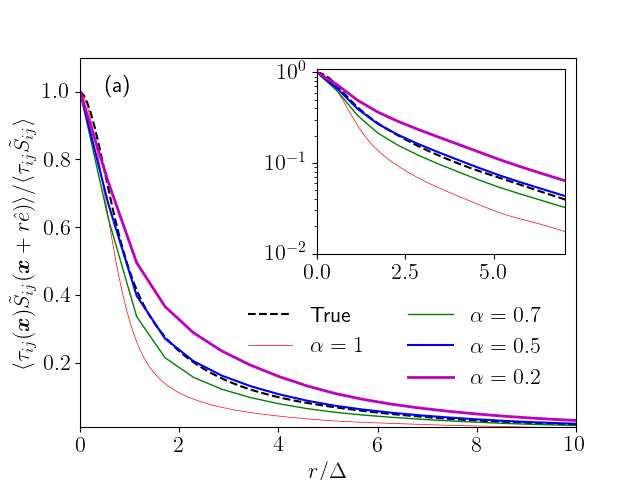}
    \includegraphics[width=0.4\textwidth]{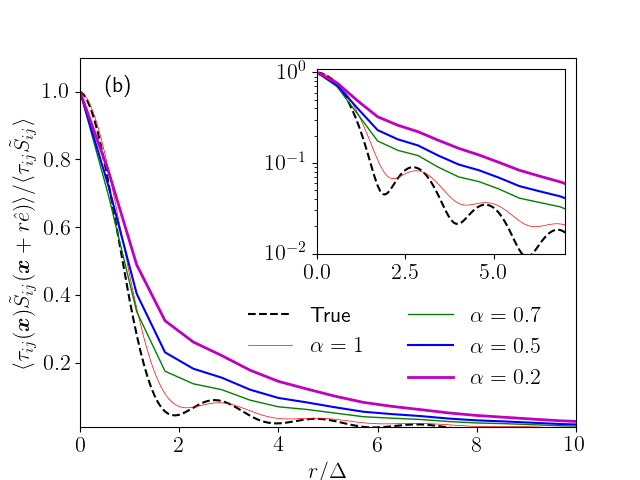}
    \caption{Two-point correlation functions between stresses and
    fractional filtered strain of different orders in HIT using two filter types, namely a Gaussian filter (a) and a spectral cutoff-filter (b).}
    \label{HIT_filters}
\end{figure*}

In Fig.~\ref{HIT_filters}(a,b)  we study the dependence of the two-point correlations with the filter type. The results using a Gaussian filter are essentially the same as those for a box-filter (we use the usual definition of $\Delta$ as summarized in \cite{Pope}). Remarkably, when using a spectral cutoff filtering (with cutoff filter equal to $k_\Delta = \pi/\Delta$ \cite{Pope}), 
the spatial correlations with the true subgrid-scale stresses decay much more rapidly than for the Gaussian and top-hat box filters. This is somewhat surprising since the spatial non-locality associated to a spectral filtering operation is more than for the box or Gaussian filters (in physical space the spectral cutoff filter's decay is slow, according to $1/r$). The oscillatory behavior is as expected. 
As a consequence, for the spectral cutoff filter the traditional local modeling appears the most appropriate. Still, as discussed in \cite{Meneveau00,eyink2009localness,aluie2009localness}, the spectral cutoff filter kernel has some undesirable features (e.g. non-positiveness in physical space kernel) and hence for the reminder we continue to focus on the physical space local box-filter. We have checked that results with the Gaussian filter lead to very similar results. 

\begin{figure}
    \centering
    \includegraphics[width=0.4\textwidth]{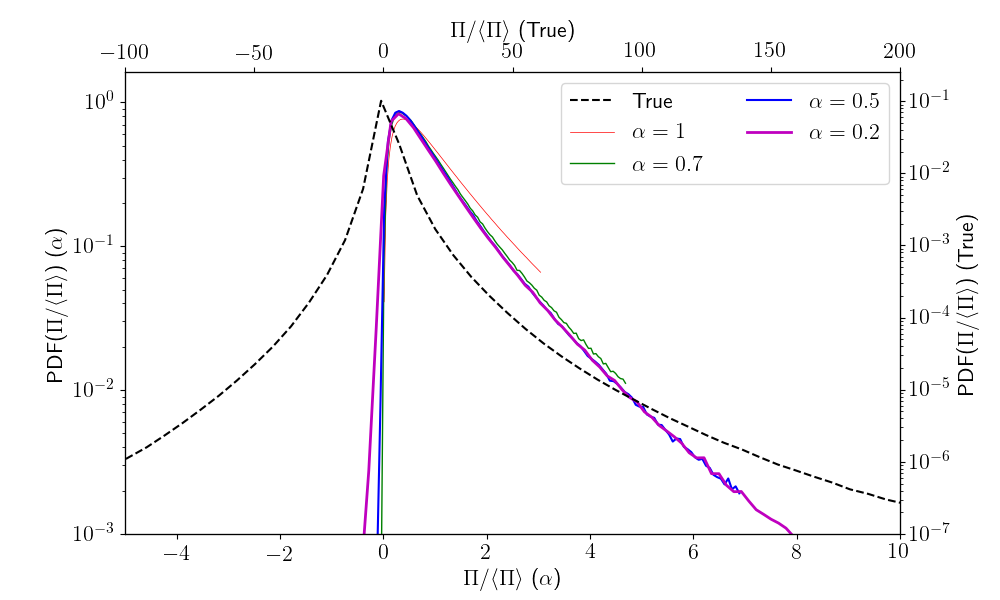}
    \caption{Probability density functions of the true and modeled subgrid dissipation $\Pi = - \tau_{ij} \tilde{S}_{ij}$ for the isotropic turbulence flow using a box filter and $\Delta=31\eta$. The bottom x-axis and left y-axis (labeled ``$\alpha$'') correspond to the modeled cases using the fractional gradient eddy-viscosity models, while the top x-axis and the right y-axis (labeled ``True'') correspond to the subgrid dissipation rate evaluated using the measured subgrid-scale stress.}
    \label{pdfs}
\end{figure}

Figure~\ref{pdfs} shows the probability density functions of the true and modeled local instantaneous subgrid-scale dissipation $\Pi = - \tau_{ij} \tilde{S}_{ij}$, for the isotropic turbulence flow using a box filter and $\Delta=31\eta$. As is well-known \cite{cerutti98}, the true distribution exhibits very long tails to both sides while the eddy-viscosity model with $\alpha = 1$ is, by definition, purely dissipative (i.e. has only positive values and much shorter tails). The results for nonlocal fractional eddy viscosity with $\alpha<1$ are almost the same as the $\alpha=1$ case but for very small probability events where $\Pi<0$ visible especially for the $\alpha = 0.2$ case.  

\subsection{Channel flow} 

\begin{figure}
    \centering
    \includegraphics[width=0.3\textwidth]{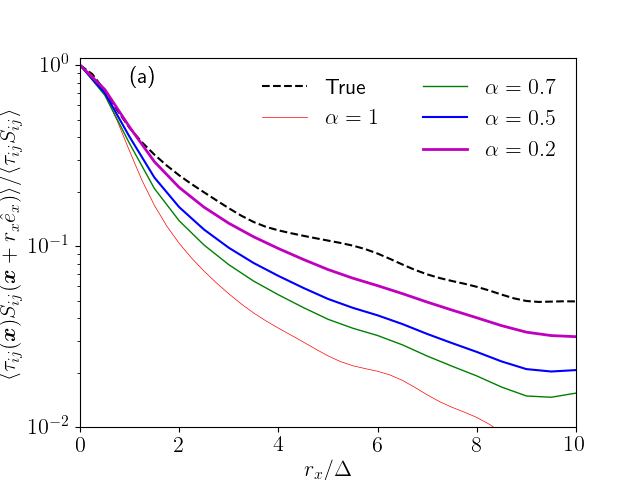}
    \includegraphics[width=0.3\textwidth]{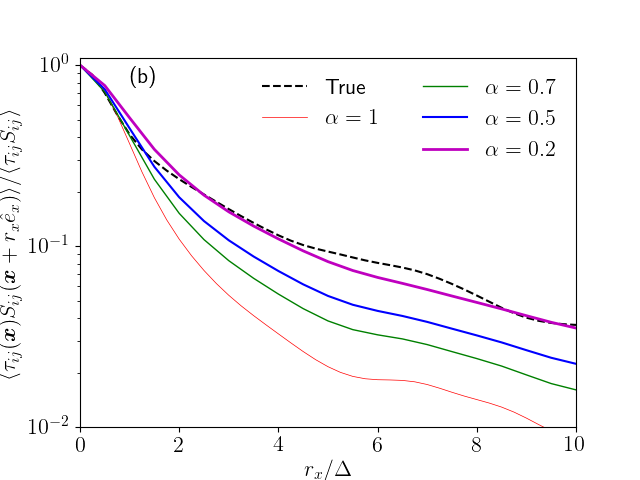}
    \includegraphics[width=0.3\textwidth]{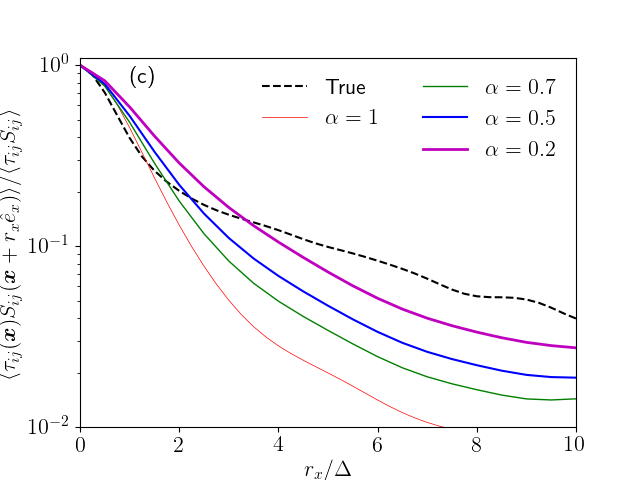}
    \caption{Streamwise two-point correlations between different
    subgrid-stresses and the filtered strain rates calculated from the
    channel flow at $\retau=1000$ data at different distances from the
    wall. Results are for top-hat filtering at $\Delta^+ = 49$ various $\alpha$ values as indicated and $R=5\Delta$. (a): $y^+=90$, (b): $y^+=260$, (c): $y^+=1000$.}
    \label{channel_top}
\end{figure}

For analysis of channel flow, we first focus on two-point correlations in the streamwise and spanwise directions. We use a top-hat filter at a scale of 
$\Delta^+ = 49$ and the filtering is performed only in the x-z horizontal directions. Data is analyzed in the logarithmic and outer regions $y^+=90, 260$ and at the centerline ($y^+=1000$ for the $\mathrm{Re}_\tau=1000$ dataset). 
Fig.~\ref{channel_top} shows the different two-point correlations along
the streamwise direction at these three different locations from the wall. Compared to the homogeneous and isotropic case,
the true correlations between the stresses and the strains along the streamwise
direction first decay rapidly and then carry on for very long distances. For distances $r>\Delta$   the local eddy-viscosity model again fails to capture these
long lasting correlations, while the introduction of non-locality via $\alpha<1$  can remedy the situation. However, there appear variations in the optimal
fractional order at  different heights with $\alpha = 0.2$ appearing to provide more realistic correlations at $y^+=260$ while at $y^+=90$ even lower values of $\alpha$ would appear to be needed. At the centerline the true correlation function has a faster initial decay and it appears that no single fractional order has the appropriate trends. 

\begin{figure}
    \centering
    \includegraphics[width=0.3\textwidth]{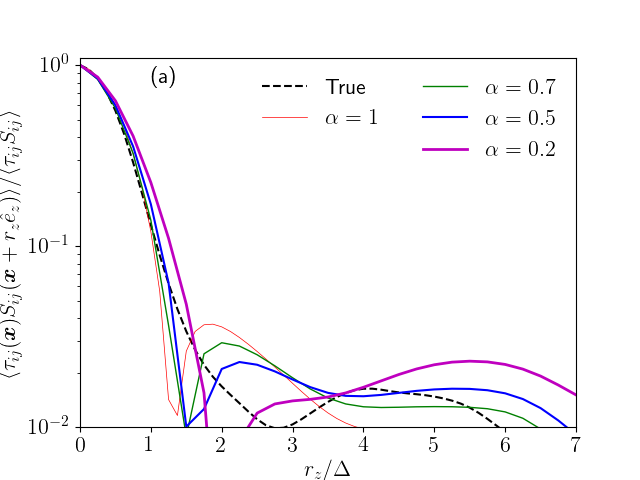}
    \includegraphics[width=0.3\textwidth]{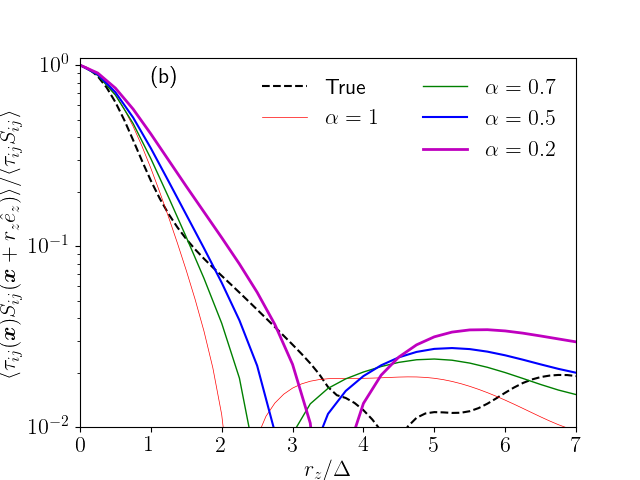}
    \includegraphics[width=0.3\textwidth]{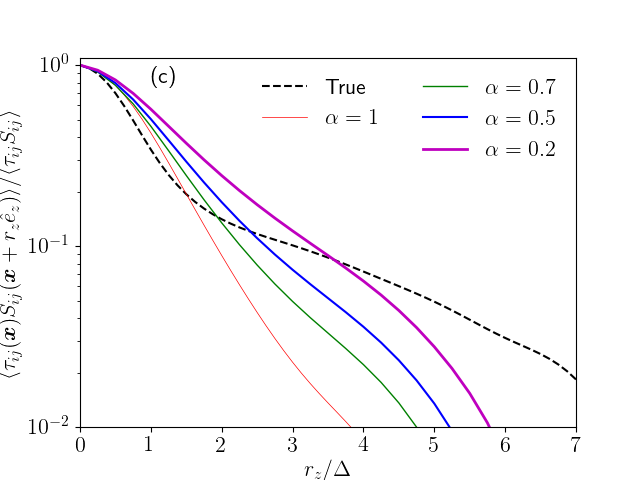}
    \caption{Spanwise two-point correlations between different
    subgrid-stresses and the filtered strain rates calculated from the
    channel flow at $\retau=1000$ data. Results are for top-hat filtering at $\Delta^+ = 49$, various $\alpha$ values as indicated and $R=5\Delta$. (a): $y^+=90$, (b): $y^+=260$ and (c): $y^+=1000$.}
    \label{channel_z}
\end{figure}

The correlations on the spanwise direction, shown in
Fig.~\ref{channel_z}, are quite different than the ones in the
streamwise direction. While a first guess would suggest that spanwise
correlations should look more similar to the homogeneous and isotropic
case, this is not the case. The behavior of the exact correlations is
not correctly captured by neither the local nor the non-local models. At the centerline, interestingly, the results for the true stress-strain rate correlations are very similar to the streamwise correlations. 

\begin{figure}
    \centering
    \includegraphics[width=0.4\textwidth]{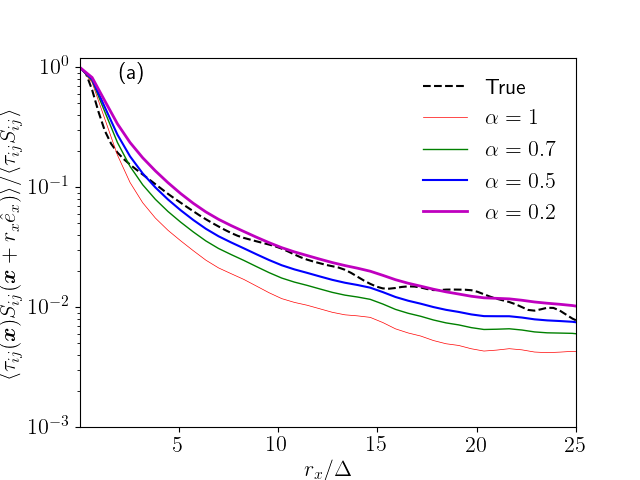}
    \includegraphics[width=0.4\textwidth]{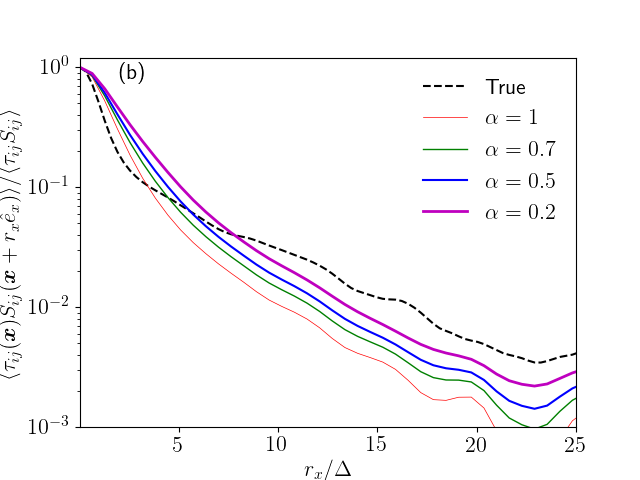}
    \caption{Streamwise two-point correlations between different
    subgrid-stresses and the filtered strain rates calculated from the
    channel flow at $\retau=5200$ data at different distances from the
    wall. Results are for top-hat filtering at $\Delta^+ = 49$, various $\alpha$ values as indicated and $R=5\Delta$. (a): $y^+=1000$, (b): $y^+=5200$.}
    \label{channel_retau}
\end{figure}

Finally, we present the two-point stress-strain rate correlation functions calculated using the channel flow data
at $\retau=5200$. Results are shown in Fig.~\ref{channel_retau} at two different heights.  The results at $y^+=1000$ are similar to the results at 
$y^+=260$ for the $\retau=1000$ dataset. We note that in outer units these two datasets are at similar heights $y/h \sim 0.2$ and 0.26 respectively. 
Also at $y^+=5200$ (the center of the
channel in this case), results are similar to the centerline results at $\retau=1000$, with the true stress-strain correlation decaying much faster at small distances and then breaking onto a very long tail. Again, none of the fractional models reproduce these trends since their decay appears to be more gradual throughout.

\section{Conclusions} 
\label{sec:conclusions}

In this paper we show that for a Large Eddy Simulation to be able to
reproduce the two-point correlations of filtered velocity fields in turbulence, 
the subgrid-stress tensor should correctly capture the two-point correlations between the filtered strain rate tensor  and the subgrid-scale stress tensor. 
This necessary statistical condition comes from the analysis of the Karman-Howarth equation for filtered velocity fields. We also generalize the derivation of \cite{Meneveau94} from homogeneous to non-homogeneous turbulence. In either case, the special importance of the correlation function 
$\langle \tau_{ki}({\bm x}) \tilde{S}_{ki}({\bm x}+{\bm r}) \rangle$ becomes apparent.

Using data from DNS of homogeneous isotropic turbulence and channel flows we show that the correlations developed by local eddy-viscosity, such as the Smagorinsky model (strictly speaking with constant eddy-viscosity), decay faster than those observed in the analysis of the true subgrid-scale stress. In order to include non-local dependencies in the model, it is argued that a convolution of the strain-rate tensor with a non-local eddy-viscosity kernel can be invoked. A mathematically compact special case of non-locality is provided by fractional differentiation. We first propose a generalization of the Caputo fractional derivative applicable to 3D problems that is amenable to vector calculus.

As a first step exploring the properties of such a modeling approach, we perform statistical a-priori testing based on DNS data from isotropic and channel flow turbulence. The analysis focuses on the behavior of predicted strain rate-stresses correlation functions that had been identified as necessary condition for LES to generate accurate predictions of two-point statistics (correlations, spectra) of filtered velocities. Different parameters are considered, such as filter size, type, wall distance (in the channel flow case) and integration radius $R$.

The main conclusion is that for many of the cases tested (filter size, type, flow), the fractional model provides more realistic predictions of the long tails in the observed two-point correlations compared to the local eddy-viscosity approach. In isotropic turbulence, a value of $\alpha \sim 0.5$ appears to provide good predictions, although we do not have a theoretical explanation for such a value. 
We note that this conclusion applies to the spatially local filters such as top-hat and Gaussian filters. For the spectral filter, it was found that the local modeling appeared appropriate. In channel flow, strong directional dependence was observed, with very strong non-locality in the streamwise direction, which is not surprising given the existence of elongated streamwise structures in this flow. The behavior in the longitudinal direction was much more local. Interestingly, at the channel centerline while the streamwise and spanwise behaviors became more similar, they differ  markedly from the behavior of isotropic turbulence. 

Clearly much more work is required before these findings can be channeled into a working subgrid model for practical applications in LES. First, the scalar prefactor $\nu_\alpha$ (the fractional eddy viscosity coefficient) must be prescribed in such a way as to enable the correct mean subgrid dissipation rate. Moreover, present results suggest that the fractional order $\alpha$ must be direction dependent in anisotropic flow, as well as depend on position (e.g. distance to the wall) in non-homogeneous flow. How to prescribe such dependencies in prognostic, general-purpose LES where one typically wishes to avoid having to use non-local information unless it arises from prognostic transport equations, is an open question. Moreover, without special treatments and accelerations, the numerical  evaluation of non-local gradients has high operations count, proportional to $R^3$ which can be quite expensive even if $R$ is restricted to $R \sim 5\Delta$. Further efforts should be directed at accelerating the evaluation of non-local operators to enable practical applications of non-local modeling. 

\vskip 0.3in

\noindent {\bf Acknowledgements:} The authors are grateful to the IDIES staff supporting JHTDB.

\noindent {\bf Funding:} Funding was provided by the AIRA (Artificial Intelligence Research Associate) program of the Defense Advance Research Projects Agency (DARPA). 

\noindent {\bf Declaration of interests:} The authors report no conflict of interest.

\noindent {\bf Author ORCIDs:}

P. Clark Di Leoni  https://orcid.org/0000-0003-3789-3466

T. A. Zaki https://orcid.org/0000-0002-1979-7748

G.E. Karniadakis  https://orcid.org/0000-0002-9713-7120

C. Meneveau https://orcid.or/0000-0001-6947-3605 
 
\noindent  {\bf Author contributions.} P.C. performed the statistical data analysis. All authors contributed to deriving theories, reaching conclusions, and writing the paper. 

\appendix

\section{Numerical technique for non-local operators}
\label{numerical_method}

We adapted an L-scheme technique \cite{Yang10} to our definition of the fractional gradient. The idea behind the method is to perform the radial integration very accurately in small discrete intervals.  In that way the singularity in the integral and in the gamma function can be  taken care of simultaneously. The method is as follows:  let $\tilde{\bm r} = \bm{r} - \bm{r}'$, $g(\tilde{\bm r}) = \frac{\partial u}{\partial r_i} (\tilde{\bm r})$, $d=3$, and use spherical coordinates in $\tilde{\bm r}$

\begin{align*}
    D^{\alpha, R}_i u(\bm{r}) &= \frac{1}{4\pi \Gamma(1-\alpha)}
    \int\displaylimits_\Omega
    \int\displaylimits^R_{0}
    g (\tilde{\bm r}) \tilde{r}^{-\alpha} d\tilde{r}
    d\Omega
    \\
    &= \frac{1}{4\pi \Gamma(1-\alpha)}
    \int\displaylimits_\Omega
    \left[
    \sum^{N-1}_{i=0} \int\displaylimits^{(i+1)h}_{ih}
    g (\tilde{\bm r}) \tilde{r}^{-\alpha} d\tilde{r}
    \right]
    d\Omega
    \\
    &\approx \frac{1}{4\pi \Gamma(1-\alpha)}
    \int\displaylimits_\Omega
    \left[
    \sum^{N-1}_{i=0} g(ih,\theta,\phi)
    \int\displaylimits^{(i+1)h}_{ih}
    \tilde{r}^{-\alpha} d\tilde{r}
    \right]
    d\Omega
    \\
    &= \frac{h^{1-\alpha}}{4\pi \Gamma(2-\alpha)}
    \sum^{N-1}_{i=0}
    \left[
    \int\displaylimits_\Omega
    g(ih,\theta,\phi)
    d\Omega
    \right]
    ((i+1)^{1-\alpha} - i^{1-\alpha})
\end{align*}
Doing this deals with both the singularity coming from the $\Gamma$ function and from the integration kernel. 

The last ingredient needed for the method is to calculate the integral over the solid angle. To do this, we first discretize the area (sphere) over where the integration takes place following the algorithm proposed by \cite{Saff97}, which generates equally spaced points on the sphere by following a spiral connecting one pole to the other. The number of integration points over each sphere, $N_i$, is chosen so that $N_i h^2 \approx 4\pi(ih)^2$, where $h$ is the desired spatial resolution. The values of the field required at all the different locations (in our case the filtered velocity gradient tensor) are obtained via trilinear spatial interpolation. The integration step $h$ is approximately equal to the grid-size of the simulations from where the data was gathered, in our case comparable to the filter size $\delta$.

\bibliographystyle{jfm}
\bibliography{references}

\end{document}